\documentclass[conference]{IEEEtran}
\IEEEoverridecommandlockouts

\usepackage{lipsum}
\usepackage[table]{xcolor}
\usepackage{multicol}
\usepackage{cite}
\usepackage{amsmath,amssymb,amsfonts}
\usepackage{algorithmic}
\usepackage{graphicx}
\usepackage{textcomp}
\usepackage{xcolor}
\usepackage{pdflscape}
\usepackage{comment}
\usepackage{acro}
\usepackage{float}
\usepackage{xcolor}
\def\BibTeX{{\rm B\kern-.05em{\sc i\kern-.025em b}\kern-.08em
    T\kern-.1667em\lower.7ex\hbox{E}\kern-.125emX}}

\DeclareAcronym{sdo}{
  short=SDOs,
  long=Standards Developing Organizations,
}
\DeclareAcronym{ai}{
  short=AI,
  long=Artificial Intelligence,
}
\DeclareAcronym{xai}{
  short=XAI,
  long=Explainable AI,
}
\DeclareAcronym{ntn}{
  short=NTNs,
  long=Non-Terrestrial networks,
}
\DeclareAcronym{itu}{
  short=ITU,
  long=International Telecommunications Union,
}
\DeclareAcronym{imt}{
  short=IMT,
  long=International Mobile Telecommunications,
}
\DeclareAcronym{embb}{
  short=eMBB,
  long=Enhanced Mobile Broadband,
}
\DeclareAcronym{urllc}{
  short=URLLC,
  long=Ultra-Reliable Low Latency Communications,
}
\DeclareAcronym{mmtc}{
  short=mMTC,
  long=Massive Machine-Type Communications,
}
\DeclareAcronym{leo}{
  short=LEO,
  long=Low Earth Orbit,
}
\DeclareAcronym{haps}{
  short=HAPS,
  long=High-Altitude Platform Systems,
}
\DeclareAcronym{uav}{
  short=UAVs,
  long=Unmanned Aerial Vehicles,
}
\DeclareAcronym{xr}{
  short=XR,
  long=Extended Reality,
}
\DeclareAcronym{ris}{
  short=RIS,
  long=Reconfigurable Intelligent Surfaces,
}
\DeclareAcronym{owc}{
  short=OWC,
  long=Optical Wireless Communications,
}
\DeclareAcronym{ue}{
  short=UE,
  long=User Equipment,
}
\DeclareAcronym{noma}{
  short=NOMA,
  long=Non-Orthogonal Multiple Access,
}
\DeclareAcronym{ofdma}{
  short=OFDMA,
  long=Orthogonal Frequency-Division Multiple Access,
}
\DeclareAcronym{tdma}{
  short=TDMA,
  long=Time Division Multiple Access,
}
\DeclareAcronym{qam}{
  short=QAM,
  long=Quadrature Amplitude Modulation,
}
\DeclareAcronym{im}{
  short=IM,
  long=Index Modulation,
}
\DeclareAcronym{oam}{
  short=OAM,
  long=Orbital Angular Momentum,
}
\DeclareAcronym{gfdm}{
  short=GFDM,
  long=Generalized Frequency Division Multiplexing,
}
\DeclareAcronym{tpc}{
  short=TPC,
  long=Turbo Product Codes,
}
\DeclareAcronym{ldpc}{
  short=LPDC,
  long=Low-Density Parity-Check,
}
\DeclareAcronym{sc-ldpc}{
  short=SC-LPDC,
  long=Spatially Coupled LDPC,
}
\DeclareAcronym{jsac}{
  short=ISAC,
  long=Joint Sensing and Communication,
}
\DeclareAcronym{5gppp}{
  short=5GPPP,
  long=5G Infrastructure Public Private Partnership,
}
\DeclareAcronym{3gpp}{
  short=3GPP,
  long=3rd Generation Partnership Project,
}
\DeclareAcronym{mmwave}{
  short=mmWave,
  long=Millimeter Wave,
}
\DeclareAcronym{thz}{
  short=THz,
  long=Terahertz,
}
\DeclareAcronym{dsa}{
  short=DSA,
  long=Dynamic Spectrum Allocation,
}
\DeclareAcronym{iot}{
  short=IoT,
  long=Internet of Thing,
}
\DeclareAcronym{fl}{
  short=FL,
  long=Federated Learning,
}
\DeclareAcronym{wsn}{
  short=WSN,
  long=Wireless Sensor Networks,
}
\DeclareAcronym{ru}{
  short=RUs,
  long=Radio Units,
}
\DeclareAcronym{du}{
  short=DUs,
  long=Distributed Units,
}
\DeclareAcronym{ran}{
  short=RAN,
  long=Radio Access Network,
}
\DeclareAcronym{los}{
  short=LOS,
  long=Line-of-Sight,
}
\DeclareAcronym{mimo}{
  short=MIMO,
  long=Multiple-Input-Multiple-Output,
}
\DeclareAcronym{sic}{
  short=SIC,
  long=Successive Interference Cancellation,
}
\DeclareAcronym{otfs}{
  short=OTFS,
  long=Orthogonal Time Frequency Space,
}
\DeclareAcronym{wbmm}{
  short=WBMM,
  long=Wavelet-Based Multicarrier Modulation,
}
\DeclareAcronym{dlt}{
  short=DLT,
  long=Distributed Ledger Technology,
}
\DeclareAcronym{ml}{
  short=ML,
  long=Machine Learning,
}

\begin{document}

\title{6G Cellular Networks: Mapping the Landscape for the IMT-2030 Framework}

\author{\IEEEauthorblockN{Ekram Hossain}
\IEEEauthorblockA{\textit{University of Manitoba}\\
ekram.hossain@umanitoba.ca
}
\and
\IEEEauthorblockN{Angelo Vera-Rivera}
\IEEEauthorblockA{\textit{University of Manitoba}\\
angelo.verarivera@umanitoba.ca
} 
\thanks{Corresponding author: Ekram Hossain (email: ekram.hossain@manitoba.ca).} 
}
\maketitle

\begin{abstract}
The IMT-2030 framework provides the vision and conceptual foundation for the next-generation of mobile broadband systems, colloquially known as Sixth-Generation (6G) cellular networks. Academic circles, industry players, and Standard Developing Organizations (SDOs) are already engaged in early standardization discussions for the system, providing key insights for future technical specifications. In this context, a structured thematic synthesis aligned with IMT-2030 is essential to inform the discussions and assist collaboration among 6G stakeholders---including scholars, professionals, regulators, and SDO officials. This review intends to offer a concise yet informative synthesis of well-established 6G literature, viewed through the IMT-2030 lens, for both specialists and generalists engaged in shaping future standards and advancing 6G research.  
\end{abstract}


\section{Introduction}
\label{Introduction}
\begin{table*}[]
\centering
\resizebox{\linewidth}{!}{
\begin{tabular}{|l|l|p{0.9cm}|p{1.4cm}|p{1.4cm}|p{1.4cm}|p{1.4cm}|p{1.4cm}|p{1.4cm}|p{1.4cm}|p{1.4cm}|}
\hline
\textbf{Survey}&\textbf{Year}&\textbf{Citation Count}&\textbf{IMT-2030 centered}&\textbf{5G Limitations} &\textbf{6G Vision}&\textbf{Use Cases}&\textbf{Architecture}&\textbf{Technology Enablers}&\textbf{Vertical Impacts}&\textbf{Research Frontier}\\
\hline
\cite{zhang2019_6g_wireless_networks} & 2019 & 1795 & \cellcolor{red!30} No & \cellcolor{red!30} No & \cellcolor{green!30} Yes & \cellcolor{green!30} Yes & \cellcolor{green!30} Yes & \cellcolor{green!30} Yes & \cellcolor{red!30} No & \cellcolor{red!30} No\\
\cite{dang2020_what_should_6g_be} & 2020 & 1330 & \cellcolor{red!30} No & \cellcolor{red!30} No & \cellcolor{green!30} Yes & \cellcolor{green!30} Yes & \cellcolor{red!30} No & \cellcolor{green!30} Yes & \cellcolor{red!30} No & \cellcolor{red!30} No\\
\cite{9144301Chowdhury} & 2020 & 1212 & \cellcolor{red!30} No & \cellcolor{green!10} Partially & \cellcolor{green!10} Partially & \cellcolor{green!30} Yes & \cellcolor{green!30} Yes & \cellcolor{green!30} Yes & \cellcolor{red!30} No & \cellcolor{red!30} No\\
\cite{9145564Akyildiz} & 2020 & 940 & \cellcolor{red!30} No & \cellcolor{red!30} No & \cellcolor{red!30} No & \cellcolor{green!30} Yes & \cellcolor{red!30} No & \cellcolor{green!30} Yes & \cellcolor{green!30} Yes & \cellcolor{green!30} Yes\\
\cite{9040431Viswanathan} & 2020 & 577 & \cellcolor{red!30} No & \cellcolor{red!30} No & \cellcolor{green!30} Yes & \cellcolor{green!30} Yes & \cellcolor{green!30} Yes & \cellcolor{green!30} Yes & \cellcolor{green!10} Partially & \cellcolor{red!30} No\\
\cite{you2020_towards6g_wireless} & 2021 & 1634 & \cellcolor{red!30} No & \cellcolor{green!30} Yes & \cellcolor{green!30} Yes & \cellcolor{green!30} Yes & \cellcolor{green!30} Yes & \cellcolor{green!30} Yes & \cellcolor{green!30} Yes & \cellcolor{green!30} Yes\\
\cite{9349624Jiang} & 2021 & 1089 & \cellcolor{red!30} No & \cellcolor{red!30} No & \cellcolor{green!30} Yes & \cellcolor{green!30} Yes & \cellcolor{red!30} No & \cellcolor{green!30} Yes & \cellcolor{red!30} No & \cellcolor{red!30} No\\
\cite{9390169Tataria} & 2021 & 968 & \cellcolor{red!30} No & \cellcolor{green!10} Partially & \cellcolor{green!30} Yes & \cellcolor{green!30} Yes & \cellcolor{green!30} Yes & \cellcolor{green!30} Yes & \cellcolor{green!10} Partially & \cellcolor{green!10} Partially\\
\cite{calvanese2021_6g_semantic} & 2021 & 368 & \cellcolor{red!30} No & \cellcolor{green!10} Partially & \cellcolor{green!30} Yes & \cellcolor{green!10} Partially & \cellcolor{red!30} No & \cellcolor{green!10} Partially & \cellcolor{red!30} No & \cellcolor{red!30} No\\
\cite{9598915Alsabah} & 2021 & 329 & \cellcolor{red!30} No & \cellcolor{green!10} Partially & \cellcolor{green!30} Yes & \cellcolor{green!10} Partially & \cellcolor{red!30} No & \cellcolor{green!30} Yes & \cellcolor{red!30} No & \cellcolor{green!30} Yes\\
\cite{Wang2023_1} & 2023 & 980 & \cellcolor{green!10} Partially & \cellcolor{green!30} Yes & \cellcolor{green!30} Yes & \cellcolor{green!30} Yes & \cellcolor{green!30} Yes & \cellcolor{green!30} Yes & \cellcolor{red!30} No & \cellcolor{green!10} Partially\\
\cite{banafaa2023_6g_mobile} & 2023 & 284 & \cellcolor{red!30} No & \cellcolor{green!30} Yes & \cellcolor{green!30} Yes & \cellcolor{green!30} Yes & \cellcolor{green!30} Yes & \cellcolor{green!30} Yes & \cellcolor{green!30} Yes & \cellcolor{green!30} Yes\\
\cite{salahdine2023_5g6g_beyond} & 2023 & 99 & \cellcolor{red!30} No & \cellcolor{green!10} Partially & \cellcolor{green!30} Yes & \cellcolor{green!30} Yes & \cellcolor{green!30} Yes & \cellcolor{green!30} Yes & \cellcolor{red!30} No & \cellcolor{green!30} Yes\\
\cite{akbar2025_challenges6g} & 2024 & 18$^*$ & \cellcolor{red!30} No & \cellcolor{red!30} No & \cellcolor{green!30} Yes & \cellcolor{green!30} Yes & \cellcolor{green!30} Yes & \cellcolor{green!30} Yes & \cellcolor{green!30} Yes & \cellcolor{green!30} Yes\\
\hline
\hline
\multicolumn{3}{|c|}{This Review} & \cellcolor{green!30} Yes & \cellcolor{green!30} Yes & \cellcolor{green!30} Yes & \cellcolor{green!30} Yes & \cellcolor{green!30} Yes & \cellcolor{green!30} Yes & \cellcolor{green!30} Yes & \cellcolor{green!30} Yes \\
\hline
\end{tabular}    
}
\caption{Thematic screening of highly cited 6G general surveys published between 2019 and 2024, following the seminal work by Saad et al.~\cite{8869705}. To the best of our knowledge, only one major survey was published in 2024, with low citation count (*). No major general surveys have been published in 2025 to date. The IMT-2030 framework was introduced in 2023.}
\label{tab:SocialApplications}
\end{table*}
%

Mobile wireless networks have evolved significantly since 1971, when Bell Laboratories introduced the cellular concept to enable spectrum reuse and boost capacity~\cite{bell1971hcmt}. Since then, five generations of cellular networks have been developed and deployed globally. As of June 2025, the \ac{3gpp}\footnote{3GPP is a consortium of telecommunications standards organizations and partner entities that develop specifications for cellular systems.} continues developing specifications for Fifth-Generation (5G)-Advanced~\cite{3gpp2023tr21918}\cite{3gpp2025tr21919}, marking the transition toward the standardization of Sixth-Generation (6G) networks~\cite{3gpp2023commit6g}. The 6G roadmap officially began in 2018 when the International Telecommunication Union (ITU)\footnote{ITU is the United Nations agency for information and communication technologies, responsible for setting global telecommunications standards.} published the report ITU-R M.2441-0, "Emerging usage trends of terrestrial IMT systems towards 2030 and beyond"\cite{itur2018usage}. This was followed by ITU-R M.2516-0, “Future Technology Trends of Terrestrial International Mobile Telecommunications Systems Towards 2030 and Beyond (IMT‑2030)”\cite{itu_r_m2516_2022} in 2022 and ITU-R M.2160-0 in 2023, which established the “Framework and Overall Objectives of the Future Development of \ac{imt}\footnote{IMT is the ITU’s generic term for mobile broadband systems, including IMT-2000, IMT-Advanced, IMT-2020, and IMT-2030—also known as 3G, 4G, 5G, and 6G, respectively.} for 2030 and Beyond,” outlining the technical goals for IMT-2030 mobile broadband systems, commonly referred to as 6G\cite{ITU2023}.

IMT-2030 defines four core design principles for 6G: \textit{Connecting the Unconnected}, \textit{Ubiquitous Intelligence}, \textit{Security and Resilience}, and \textit{Sustainability}. It envisions a globally connected platform that supports massive user densities with data rates reaching hundreds of Gbps, near real-time latency, and near-perfect availability. 6G aims to overcome 5G’s limitations by advancing transmission capacity, coverage, reliability, security, intelligence, and energy efficiency. It opens the door to new Internet applications and business models across industry verticals\footnote{In economics and market theory, verticals refer to specific economic or industry sectors representing target market segments.} with potential to transform how we manage human life on our planet. 

The 6G literature is expanding rapidly, with several studies attempting comprehensive reviews of the topic. The first major survey by Saad et al.\cite{8869705} in 2019 presented an early vision for 6G systems and was followed by numerous peer-reviewed articles, white papers, and technical reports. Prior to the release of the IMT-2030 framework, most prominent surveys offered broad overviews or focused on specific research areas. These early works typically proposed their particular vision and use cases for the system, identified technology enablers, and outlined research agendas. However, they often gave limited attention to network architecture, management frameworks, societal implications, and the research frontier for enablers. Examples include~\cite{zhang2019_6g_wireless_networks, dang2020_what_should_6g_be, 9144301Chowdhury, 9145564Akyildiz, 9040431Viswanathan, you2020_towards6g_wireless, 9349624Jiang, 9390169Tataria, calvanese2021_6g_semantic, 9598915Alsabah}. The release of the IMT-2030 framework in 2023 provided a unified vision with defined use cases and technical requirements. Nonetheless, most post-2023 surveys still overlook its foundational principles for organizing 6G research. For instance,~\cite{Wang2023_1, banafaa2023_6g_mobile, salahdine2023_5g6g_beyond, akbar2025_challenges6g} omit or only partially address IMT-2030 in their analyses of 6G networks.

As we approach 6G standardization, 3GPP is hosting summits and workshops in collaboration with academia, industry, and Standards Development Organizations (SDOs)\cite{3gpp_browse_2024}. These events are promoting early discussions guided by the IMT-2030 framework\cite{5gAmericas_IMT2030_2024}, providing key insights that will shape future 6G technical specifications. In this context, a structured thematic synthesis aligned with IMT-2030 is essential to inform ongoing discussions, identify research directions, and assist collaboration among 6G stakeholders. Our main contributions are:
\begin{enumerate}
    \item A thematic synthesis of well-established 6G research, organized from the perspective of the IMT-2030 framework.     
    \item Identification of open problems and challenges in key 6G research areas. 
    \item A structured 6G knowledge base that support early standardization discussions, facilitates stakeholder collaboration, and guide future standards-informed research.
\end{enumerate}
\section{Methodology and Organization}
\begin{figure}
    \centering
    \includegraphics[width=\linewidth]{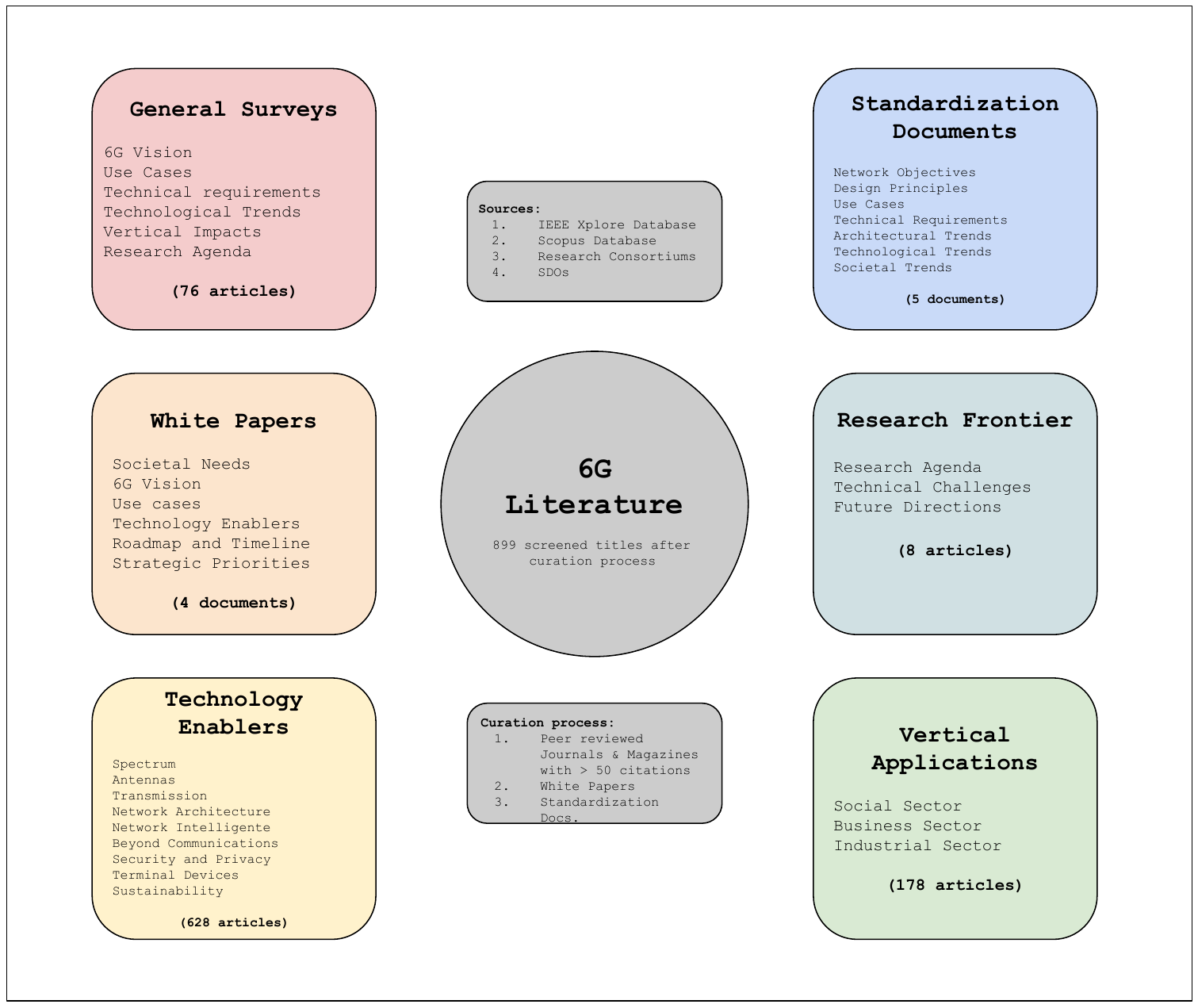}
    \caption{Map of the literature selection categorizing publications by type and thematic dimension, including general surveys, white papers, technology enablers, vertical applications, research frontiers, and standardization documents.}
    \label{fig:LiteratureMap}
\end{figure}
%

\subsection{Methodology}
This review adopts a semi-systematic methodology to explore and consolidate the 6G technological landscape, aligning with the IMT-2030 framework. The approach is well-suited for emerging and dynamic fields like 6G, enabling the identification, analysis, and synthesis of key contributions across core thematic areas~\cite{snyder2019literature, tranfield2003methodology, vom2015conducting}. The review is organized around five thematic dimensions: (1) 6G Vision, (2) Use Cases, Performance Requirements, and Architectural Trends (3) Enabling Technologies, (4) Impact on Vertical Sectors, and (5) the 6G Research Frontier. These core themes follow an evolution-oriented structure that mirrors the transition from 5G to 6G, in line with the design principles, use cases, technical capabilities, and technological trends outlined by the ITU for IMT-2030. The goal is to provide a concise yet informative synthesis for 6G stakeholders---generalists and specialists---engaged in shaping future technical standards and advancing research efforts in the field.

To ensure the inclusion of well-established contributions, we selected highly cited academic literature, seminal white papers, and official documentation from SDOs aligned with the IMT-2030 vision and objectives. We prioritize highly-visible articles to capture research directions receiving the most attention. However, we acknowledge this approach may overlook high-quality recent studies with limited visibility. Selecting appropriate scientific databases for literature screening was challenging given the numerous options covering the field. However, we learned that most relevant 6G literature is primarily indexed in IEEE Xplore and Scopus. We screened English-language articles published between 2019 and 2025 in peer-reviewed journals, magazines, and conference proceedings with a minimum citation count of 50. We also included relevant early standard documents from the ITU and seminal white papers from major research consortia, such as the 6G Flagship project~\cite{6gflagship2025}, Hexa-X~\cite{hexax2021d1.1}, 5G Americas~\cite{5gamericas2024imt2030}, and the Asia-Pacific Telecommunity (APT)~\cite{apt2022imt2030}. Search terms used Boolean combinations of: “5G and Beyond,” “6G,” “IMT-2030,” “standardization,” “use cases,” “technical requirements,” “enabling technologies,” “applications,” “research challenges,” “research directions,” and “research agenda.” 

\begin{figure}
    \centering
    \includegraphics[width=\linewidth]{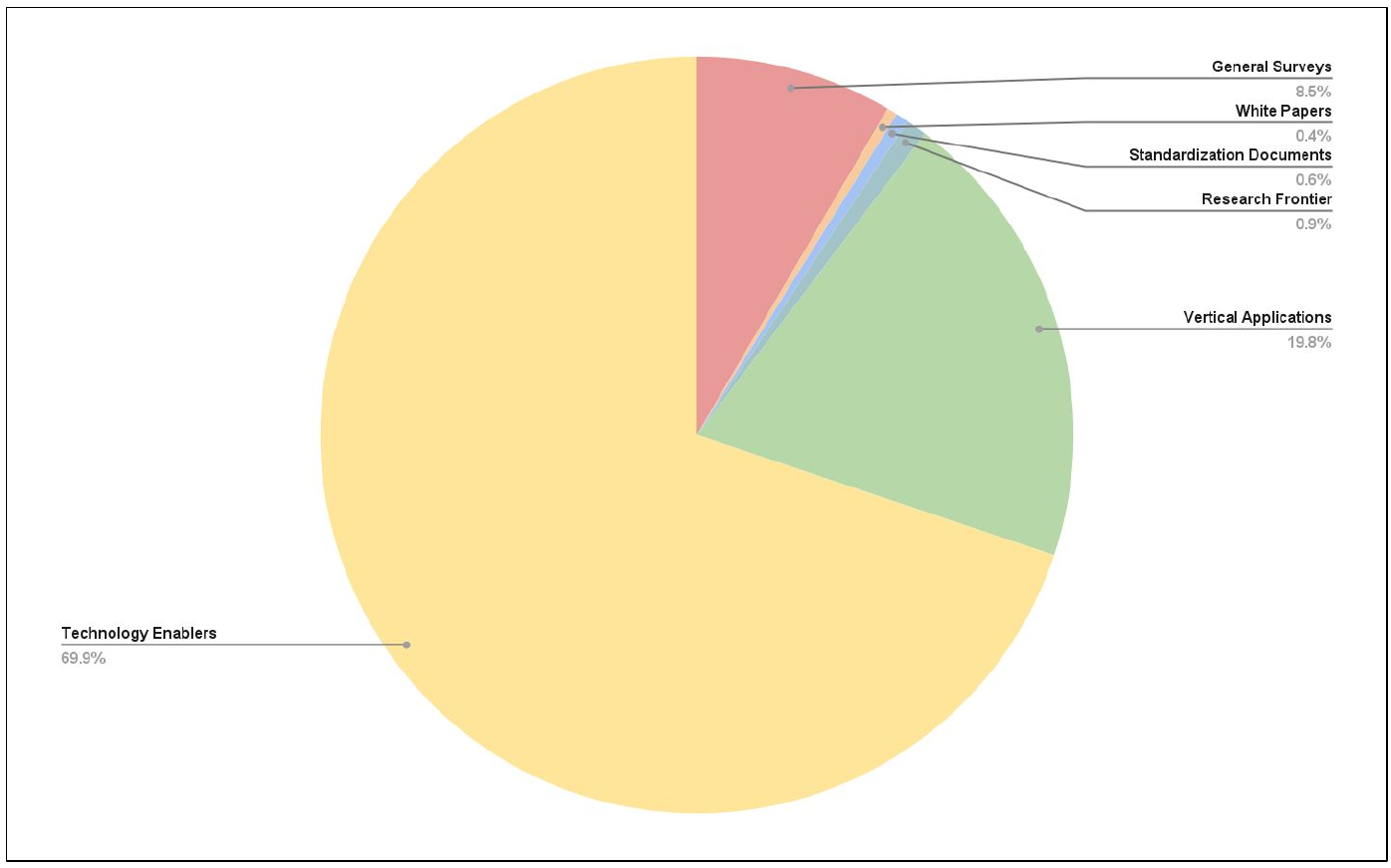}
    \caption{Percentage distribution of the screened literature by thematic dimension. Most reviewed publications focus on Technology Enablers (69.9\%), followed by Vertical Applications (19.8\%) and General Surveys (8.5\%). A small share of the literature corresponds to the Research Frontier for the system (0.9\%), Standardization Documents (0.6\%), and 6G White Papers (0.4\%).}
    \label{LiteratureComposition}
\end{figure}

A literature map organized by thematic dimension along with the percentage distribution of the contributions are shown in Figures~\ref{fig:LiteratureMap} and~\ref{LiteratureComposition}, respectively. The technology enablers theme represents the largest share of screened literature (628 articles). Table~\ref{tab:TechnologyEnablers} shows the research areas, key directions, and the corresponding percentage breakdown under this theme. For transparency, the full list of screened publications is available in our open-access archive repository~\cite{6gresearchrepo2025}. Future syntheses efforts could extend this review by employing stricter methodologies such as systematic or meta analysis to enable a more quantitative evaluation of each thematic dimension within the IMT-2030 framework.

\subsection{Organization}
The article is structured as follows. Section III examines key 5G limitations and presents the unified vision for 6G anchored in the IMT-2030 conceptual principles. Section IV outlines use cases, technical requirements, and architectural trends for the system. Section V explores technology enablers across scientific domains supporting 6G capabilities. Section VI discusses the impact and implications of 6G in the social, business, and industrial sectors. Section VII identifies the research frontier and future research directions beyond 6G. The article ends with a conclusions section.
\begin{table}[]
\centering
\resizebox{\linewidth}{!}{
\begin{tabular}{|l|l|r|}
\hline
\textbf{Research Area} & \textbf{Research Direction}  & \textbf{Percentage} \\
\hline
Spectrum &  New transmission frequencies & 13.22\% \\
 &  Spectrum management & 0.64\% \\
Antennas &  Antenna arrays \& meta-surfaces & 20.22\% \\
 &  New antenna types and designs & 1.43\%  \\
Transmission & New multiple access techniques & 4.78\% \\
& New radio architectures & 4.14\% \\
& Modern modulation and codign techniques & 1.27\% \\
Network Architecture & Heterogeneous network integration  & 14.49\% \\
& Distributed network orchestration & 4.78\% \\
& Distributed computing infrastructure & 4.46\% \\
Network Intelligence &  Integration of modern AI models & 13.54\% \\
Beyond Communications & Joint sensing, communication, and computing  & 6.69\% \\
& Semantic communications & 1.27\% \\
Security and Privacy & Distributed security and privacy models & 4.46\% \\
Terminal Devices & New device materials, types, and designs & 1.11\% \\
Sustainability & Energy-aware communication protocols  & 1.91\% \\
 & Green communication infrastructure & 1.59\% \\
\hline
\end{tabular}    
}
\caption{Research areas, key directions, and the percentage breakdown of the screened literature under the Technology Enablers theme (628 titles). This theme represents the largest share of available 6G literature.}
\label{tab:TechnologyEnablers}
\end{table}

%
%

\section{6G Panoramic Vision: Expanding the Scope of Connectivity}
\label{6GVision}
\begin{figure*}
    \centering
    \includegraphics[width=0.65\linewidth]{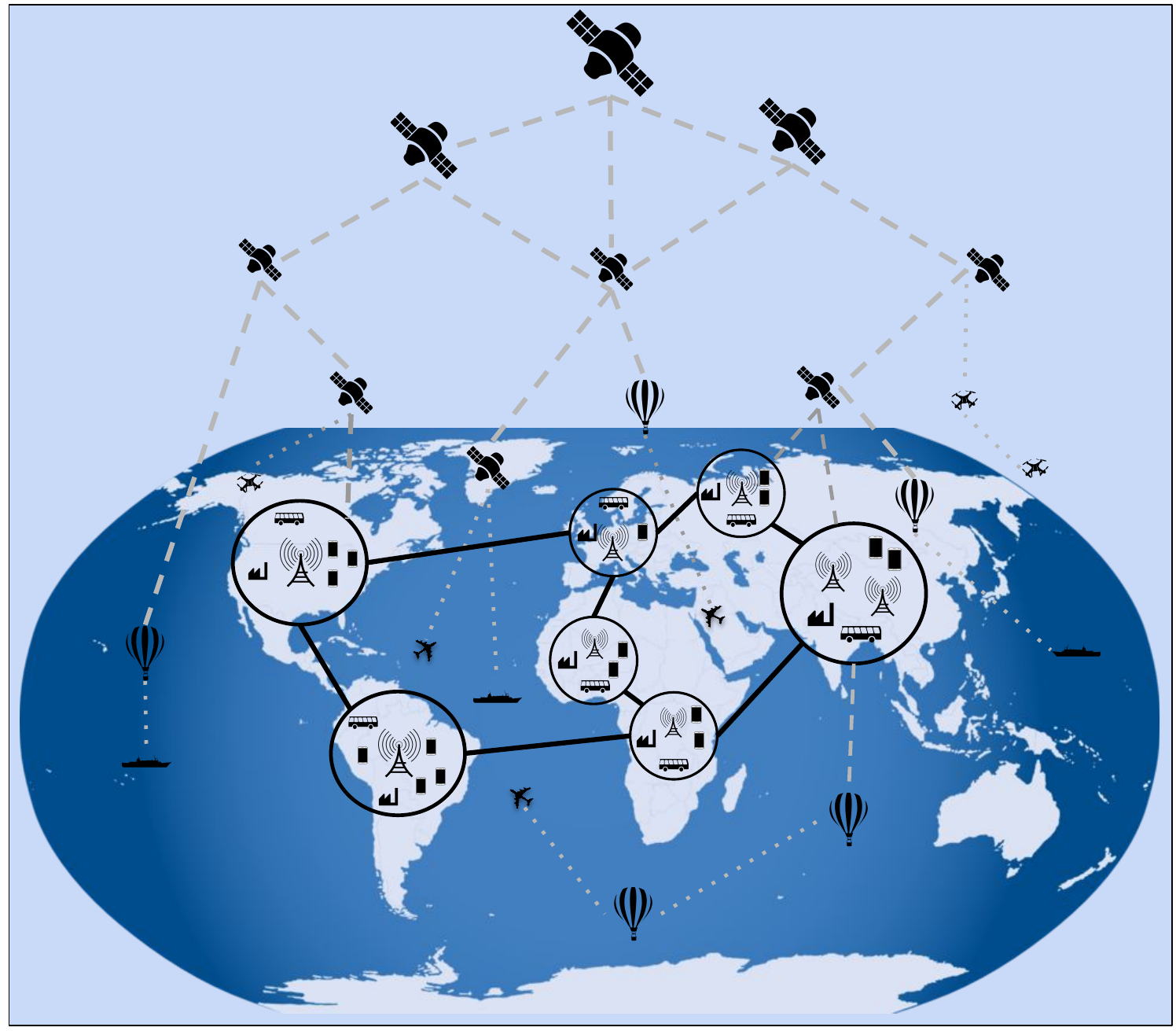}
    \caption{Vision of global 6G coverage enabled by NTNs illustrating the integration of LEO satellites, HAPS, and UAVs with terrestrial infrastructure to ensure continuous, resilient, and high-performance connectivity across land, sea, air, and space. The solid lines represent ground or undersea connections. Non-solid lines represent air connections.}
    \label{Figure_1}
\end{figure*}
%
During the last three decades, mobile broadband systems have evolved from 3G's basic Internet access to 4G's high-speed streaming and interactive apps, and now to 5G's support for intelligent transportation, industrial automation, and smart urban infrastructure. 5G significantly advanced mobile broadband technologies, delivering significant performance improvements over previous generations~\cite{Gupta2015}. However, it still faces critical limitations to meet increasing connectivity demands from users, business models, and industrial applications~\cite{Moussaoui2022}. These demands---including global coverage, precision positioning, ubiquitous intelligence, and carbon neutrality---extend beyond the traditional \textit{``more bandwidth, less latency"} conversation, exceeding 5G's capabilities. 5G networks incorporate \ac{mmwave} spectrum (i.e., 24–100 GHz) to achieve ultra-high data rates and low latency. However, \ac{mmwave} signals exhibit major propagation challenges, limiting coverage and Quality of Service (QoS) \cite{Al-Ogaili2016}. They suffer from high path loss due to increased attenuation at higher frequencies. Also, their short wavelengths---in the order of the millimeters---limit diffraction and make them more susceptible to blockage by buildings, trees, and other dense objects. Additionally, they are prone to scattering and experience significant absorption when penetrating materials with high dielectric constants, such as concrete and insulated glass \cite{Souza2022} \cite{Banday2018}.

5G is a complex system requiring significant real-time decision-making. While \ac{ai} is critical for optimizing and managing 5G networks~\cite{Morocho2018}, it was not formally included in the standard. Its full potential remains underutilized due to limited edge and fog computing resources, security and privacy concerns during training, and the absence of a clear system-level vision for \ac{ai} at the time of defining the standards. Additionally, 5G infrastructure demands more energy than previous generations to support massive MIMO, beamforming, always-on components, and legacy system integration~\cite{Tan2022}, making it difficult to align with carbon neutrality goals. Navigating 5G limitations provide key insights to envision the path towards future mobile broadband networks.

6G aspires to go beyond connectivity, enabling intelligent and secure ecosystems where everything is interconnected in near real-time. While 6G might seem like the next iteration in mobile broadband evolution, the system is far more than just communications. 6G envisions global connectivity, immerse experiences, high-precision positioning, and enhanced contextual awareness \cite{Wang2023_1}. It will enable a seamless fusion of physical and digital realms, unlocking a new paradigm of enriched and surreal experiences for users. On a colorful note, the idea of a surreal coexistence between the physical and digital worlds was introduced by Neal Stephenson in his 1992 novel Snow Crash\cite{Stephenson1992}. He writes, “In the lingo, this imaginary place is known as the Metaverse. Hiro spends a lot of time in the Metaverse. It beats the sht out of the U-Stor-It.”  In Hiro's Metaverse, individuals can immerse themselves in a fictional 3D universe that allows them to escape their dystopian reality. With the vision of 6G, such a world may move from fiction to reality, enabling revolutionary Internet applications and hyper-connected experiences\cite{Wang2023_2}.

\subsection{Connecting the Unconnected}
As shown in Figure~\ref{Figure_1}, 6G envisions global connectivity, delivering seamless, high-capacity service across land, sea, air, and space. The system will integrate terrestrial and non-terrestrial networks (NTNs), including LEO satellites, HAPS, and UAVs to serve users in urban, rural, airborne, and maritime environments \cite{Araniti2022}. Transitions between these nodes will be smooth, ensuring uninterrupted mobility and QoS. Global coverage will be supported by an extended transmission frequency range, moving from mmWave into the THz (100 GHz to 10 THz) and visible light bands. While these frequencies enable ultra-fast data rates (up to a few hundreds of Gbps), they also face significant propagation challenges \cite{Shafie2023}. These limitations may be mitigated by advanced network integration models, modern antenna arrays and reconfigurable meta-surfaces, intelligent beamforming, and dynamic spectrum management strategies, enabling high-performance communication across the globe~\cite{Wan2021TerahertzMIMO}.

\subsection{Ubiquitous Intelligence}
6G will integrate AI with sensing capabilities and real-time data processing across the radio, transport, and core networks to support intelligent, adaptive, and self-optimizing network functions and services \cite{Chen2020Roadmap6G}. 6G ecosystems will support extensive data collection through massive sensor networks, enabling both intelligent network operation and context-aware applications \cite{Mahmood2021MTC6G}. The integration will give connected entities—--such as robots, homes, industrial systems, and urban infrastructure---environmental awareness and situational intelligence. It will also enable virtual replicas of physical systems---i.e., digital twins---that can interact with the real world~\cite{Lin2023}. These twins will open the door for the Artificial Intelligence Generated Content (AIGC)-based economy, creating markets for AI-generated assets and services on 6G platforms~\cite{Li2024}.

\subsection{Security and Resilience}
Leveraging network data for AI-native 6G introduces significant privacy and security challenges~\cite{siriwardhana2021ai6g}. Communication platforms handle vast volumes of sensitive subscriber information, including location, behavioral patterns, and usage history. When this data is used to train AI models to optimize network functions, it becomes susceptible to malicious attacks and accidental breaches, putting user privacy at risk~\cite{munir2022security}. The tension between 6G's intelligence vision and fundamental data privacy rights can be addressed through privacy-preserving AI models and blockchain-based zero-trust data management frameworks~\cite{Porambage2021_6GSecurity}. Integrating blockchain with AI in 6G can ensure transparent governance, accountable decision-making, and traceable behavior, enhancing the network’s ability to operate securely in high-risk environments~\cite{Hewa2020}.

\subsection{Sustainability}
In 6G platforms, AI components will be integrated with cloud, edge, and fog computing infrastructure to enable data processing and decision-making capabilities near end users~\cite{kong2022edge_iot_survey}. Given the high computational demands of AI, local-level processing is essential for latency-sensitive applications~\cite{Ahammed2022}. However, supporting extensive computing infrastructure requires significant amounts of energy, raising concerns the environmental impact of 6G~\cite{jiang2023networking}. To address this, 6G development must prioritize low-complexity communication and intelligence algorithms, energy-efficient antenna and sensor materials, the use of renewable energy sources, and sustainable deployment and operation practices to minimize the system’s overall carbon footprint~\cite{huang2020green6g,Hu2021InNetwork,Zhang2022GreenerRAN}.
\section{Technical Definitions: Understanding the Demands of the IMT-2030 Framework}
\label{TechnicalRequirements}
ITU-R M.2160-0~\cite{ITU2023} outlines six use cases and fifteen technical requirements for IMT-2030, each directly aligned with the system's core design principles. We discuss use cases, technical requirements, and architectural trends in the following subsections.
\begin{table*}[]
\centering
\resizebox{\linewidth}{!}{
    \begin{tabular}{|p{4cm}|p{4cm}|p{3.8cm}|p{4cm}|}
        \hline
        \textbf{Technical Requirement} & \textbf{IMT-2020 (5G)} & \textbf{IMT-2030 (6G)} & \textbf{6G $>$ 5G} \\
        \hline
        Peak Data Rate & Up to 20 Gbps & Up to 200 Gbps & 10x higher \\
        User Experienced Data Rate & $\sim$ 100 Mbps & Between 300 and 500 Mbps & 3 to 5x higher \\
        Spectrum Efficiency & $\sim$ 30 bps per Hz & Between 45 and 90 bps per Hz & 1.5 to 3x higher \\
        Area Traffic Capacity & $\sim$ 10 Mbps per m$^2$ & Between 30 and 50 Mbps per m$^2$ & 3 to 5x higher  \\
        Connection Density & Up to 10$^6$ devices per km$^2$ & Up to 10$^8$ devices per km$^2$ & 2 orders of magnitudes higher \\
        Mobility & Up to 500 kmph & Between 500 and 1000 kmph & 2x higher (upper value)\\
        Latency & Down to 1 ms & Between 0.1 and 1 ms & 10x faster (lower value)\\
        Reliability & 1x10$^{-5}$ & Between 1x10$^{-7}$ and 1x10$^{-9}$ & 2 to 4 orders of magnitude better \\
        Coverage & Limited air and ground & Space orbit, air, ground, and sea & Global $>$ Non-global \\
        Positioning Accuracy & $\sim$ 10 m & Between 1 and 10 cm & 100x higher (worst case scenario) \\
        Sensing-Related capabilities & N/A & Candidate metrics: detection rate, accuracy, and resolution & Novel requirement defined for IMT-2030  \\
        Applicable-AI Related Capabilities & N/A & Metrics not yet defined & Novel requirement defined for IMT-2030 \\
        Security and Resilience & N/A & Metrics not yet defined & Novel requirement defined for IMT-2030 \\
        Sustainability & Up to 10 Mbps per Joule & Up to 1 Gbps per Joule & 100x better \\
        Interoperability & Primarily integrated with 4G and WiFi & Seamless integration with most legacy systems & Broader range of interoperable technologies \\
        \hline
    \end{tabular}
}
\caption{Side-by-side comparison of IMT-2020 (5G) and IMT-2030 (6G) across the fifteen technical capabilities defined in ITU-R M.2160-0~\cite{ITU2023}. Performance targets for IMT-2020 are based on ITU-R M.2083-0~\cite{ITU2015}.}
\label{Table_4}
\end{table*}   
\subsection{Use Cases and Technical Requirements}
IMT-2030 expands the three IMT-2020 (5G) use cases---\ac{embb}, \ac{urllc}, and \ac{mmtc}~\cite{ITU2015}---into six new network scenarios: \textit{Immersive Communication}, \textit{AI and Communication}, \textit{Hyper-Reliable and Low-Latency Communication}, \textit{Ubiquitous Connectivity}, \textit{Massive Communication}, and \textit{Integrated Sensing and Communication}~\cite{ITU2023}. These use cases are the starting point for defining technical requirements, identifying enabling technologies, shaping the architecture for the system, and guiding future standards. Technical requirements are presented in Table~\ref{Table_4}, and a detailed discussion of IMT-2030 use cases follows.

\subsubsection{Immersive Communication} extends the eMBB use case to enable 3D immersive applications that blend physical and digital environments. It targets ultra-high data rates in the hundreds of Gbps, sub-millisecond latency, extreme reliability, massive connectivity, haptic feedback, and seamless AI integration to support rich, real-time interactions. This use case will leverage \ac{xr}, holographic tele-presence, and multi-sensory tactile interfaces to power applications such as virtual meetings, accessible content delivery, and lifelike social media~\cite{Shen2023Immersive6G}. These services require robust infrastructure capable of high-throughput processing in challenging wireless conditions. Key technology enablers include AI, distributed computing, THz communications, ultra-massive MIMO, and blockchain for secure and efficient interactions. \textit{Immersive Communication} is expected to impact verticals such as healthcare, education, and remote work by improving accessibility and user experience~\cite{AllamJones2020_6Gsmartcities}.

\subsubsection{Hyper-Reliable and Low-Latency Communication} enhances the URLLC use case with stricter performance targets~\cite{siddiqui2023urllc}. It requires sub-millisecond latency, hundred of Gbps data rates, and extreme reliability to support mission-critical applications where performance failures could compromise user safety. These applications include emergency response, tele-medicine, infrastructure monitoring, and autonomous driving~\cite{khan2022urllc_embb_iiot,Ranjha2022URLLC}. Achieving such performance is challenging---especially in dense environments---due to interference, multipath fading, and the propagation limitations of high frequency signals. Security is also critical, given sensitive data might be transmitted over potentially vulnerable channels. Key enablers for this use case include advanced modulation and coding, ultra-massive MIMO, NTNs, distributed computing, lightweight cryptographic techniques, and AI-driven control.

\subsubsection{Massive Communication} extends the mMTC use case to enable large-scale wireless sensor networks. It targets ultra-high device density in the hundreds of millions per km$^2$, wide coverage, sub-millisecond latency, high reliability, and low energy consumption. Deploying massive device environments is challenging due to the need for extensive and energy-intensive transmission infrastructure~\cite{Mahmood2021MTC6G,Yadav2024EnablingMIoT6G}. This use case will leverage technologies such as ultra-massive MIMO, reconfigurable intelligent surfaces (RIS), AI, dynamic spectrum management, blockchain, and green energy sources to meet its demands. \textit{Massive Communication} is vital for IoT-driven sectors, supporting near real-time data exchange, intelligent control, and automation across numerous verticals, including smart cities, factories, transportation, utilities, healthcare, and retail.

\subsubsection{Ubiquitous Connectivity} is a new IMT-2030 use case focused on delivering seamless connectivity across space, air, land, and sea, including remote and under-served regions \cite{Chen2020Vision6G}. 6G will extend 5G’s coverage by integrating terrestrial and NTNs to enable omnipresent, intelligent, and context-aware communication~\cite{giordani2021ntn6g}. Key requirements include global coverage, massive connectivity, ultra-low latency, high reliability, high mobility, interoperability, and low energy consumption. Enabling technologies include LEO satellites, HAPS, network integration models, ultra-massive MIMO, RIS, optical wireless communication (OWC), and distributed computing~\cite{cui2022sagin,xiao2021antenna}. This use case supports ultra-reliable communication anytime, anywhere---crucial for logistics, transportation, public safety, and emergency response in areas beyond the reach of previous mobile broadband generations.

\subsubsection{Artificial Intelligence and Communication} is also a new IMT-2030 use case seeking to natively embed intelligence into the network functions and services~\cite{AlQuraan2023EdgeNative}. Native intelligence involves collecting, processing, and using data for AI training and operation stages across radio, transport, and core networks~\cite{zhang2019_6g_vision}. Communication requirements include sub-millisecond latency, high reliability, massive device connectivity with sensing capabilities, and extensive computing resources. Key enablers include \ac{fl}, \ac{xai}, distributed computing, and energy-efficient infrastructure~\cite{yang2021federated_learning_6g,Guo2019XAI6G,AlAnsi2021}. This use case supports self-organizing, self-optimizing, and self-healing networks for autonomous management across multiple network layers and domains.

\subsubsection{Integrated Sensing and Communication} is the final IMT-2030 use case, designed to enhance multi-dimensional sensing and contextual awareness for the network~\cite{liu2022isas}. It targets both connected and unconnected objects, capturing spatial layout, objects motion (e.g., position, velocity, trajectory, rotation), human mood and behavior (e.g., posture, gestures), and environmental conditions (e.g., rain, pollution)~\cite{Tan2021ISAC}. Communication requirements include sub-millisecond latency, extreme reliability, extended coverage, high-resolution sensing, extensive computing resources, security assurances, and energy-efficient infrastructure. Key enablers include \ac{wsn}, ultra-massive MIMO, RIS, AI, digital twins, distributed computing, blockchain, and low-power sensors and actuators. This use case is critical for context-aware services in transportation, healthcare, environmental monitoring, and public safety~\cite{GonzalezPrelcic2024ISC}.
%
%
\subsection{Architectural Trends}
\label{NetworkArchitecture}
The architecture for 6G must provide a well-defined structural framework for organizing hardware components, connections, protocols, and data planes. The \ac{5gppp}\footnote{5GPPP is a joint initiative between the European Commission and the European ICT sector focused on next-generation communication solutions. 5GPPP contributes technical studies, reports and proposals to ITU Study Groups} through their Architecture Working Group outlines eight guiding principles for the 6G architecture: \textit{Exposure of Capabilities}, \textit{AI for Full Automation}, \textit{Flexibility to Different Topologies}, \textit{Scalability}, \textit{Resilience and Availability}, \textit{Service-Based Exposed Interfaces}, \textit{Separation of Network Function Concerns}, and \textit{Network Simplification}~\cite{5GPPP2022}.

The \textit{Exposure of Capabilities} principle ensures that the 6G architecture expose performance metrics (e.g., latency, throughput, localization, sensing) to user applications and network functions for predictive orchestration. The \textit{AI for Full Automation} pillar emphasizes autonomous network management through AI without human intervention. The \textit{Flexibility to Different Topologies} idea highlights seamless integration of various network types (e.g., public, private, ad-hoc, etc.) under dynamic traffic and spectrum conditions. Scalability requires support for deployments of varying sizes, enabling elastic adaptation to user and resource demands. The \textit{Resilience and Availability} postulate demands robust infrastructure resistant to failures from attacks, accidents, or natural disasters. The \textit{Service-Based Interfaces} concept call for cloud-native, standardized, service-oriented interfaces across network layers to ensure smooth data and function interaction. Separation of Concerns promotes modular network functions with clear responsibilities and minimal dependencies. Lastly, the \textit{Network Simplification} philosophy aims to reduce architectural complexity through fewer configuration parameters and streamlined interfaces.

\begin{figure}[h]
    \centering
    \includegraphics[width=\linewidth]{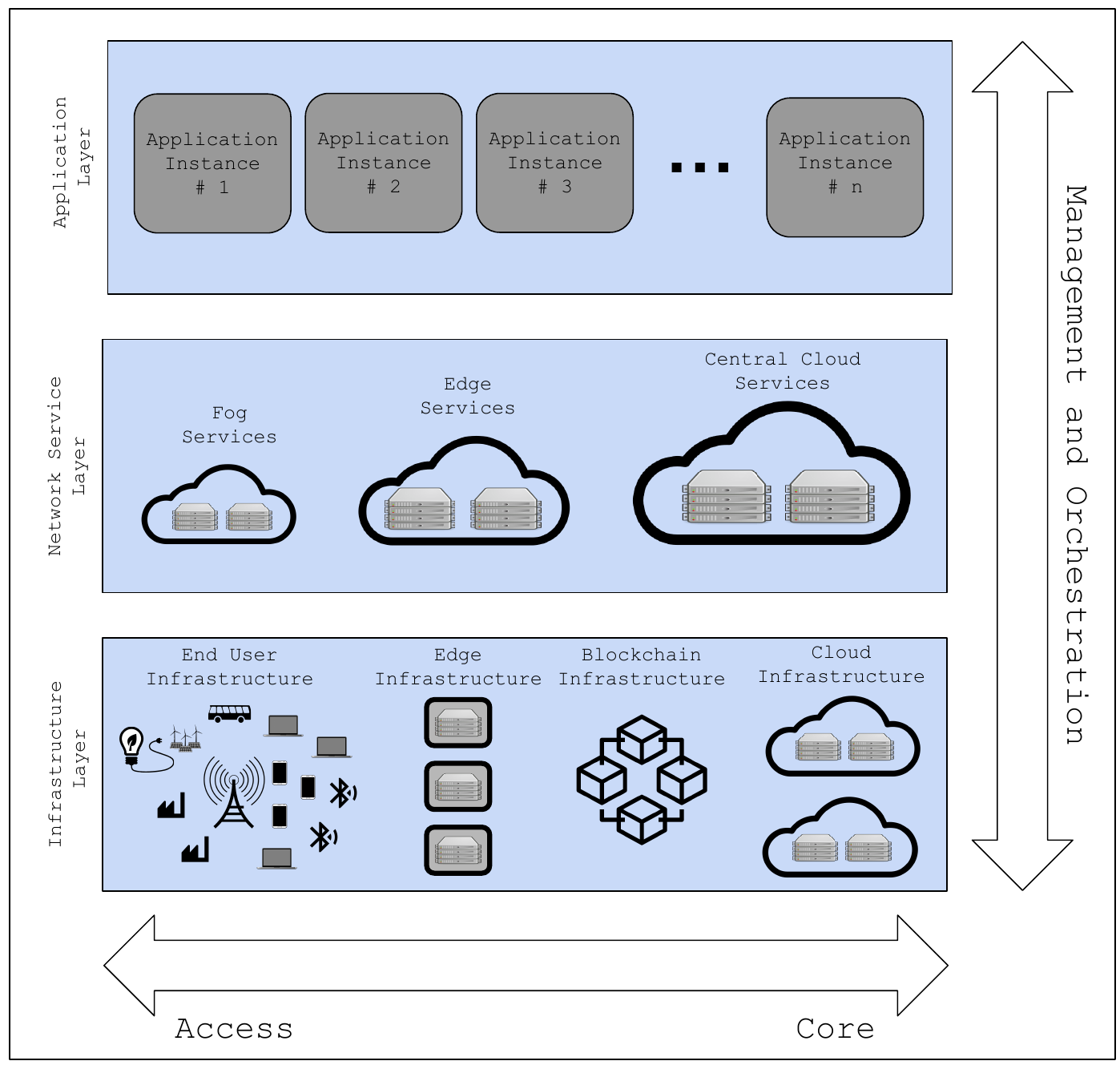}
    \caption{Layered high-level architecture for 6G networks illustrating the interplay between infrastructure (including terrestrial, aerial, and satellite components), network services (such as AI-native orchestration, slicing, and trust mechanisms), and application layers.}
    \label{6GArchitecture_2}
\end{figure}

The architecture is preliminarily organized into three high-level layers: \textit{Infrastructure}, \textit{Network Services}, and \textit{Applications}.
\begin{enumerate}
    \item The \textit{Infrastructure Layer} comprises the radio, transport, and core networks, including components such as RUs, DUs, switches, routers, data centers, and both edge and cloud computing infrastructure. In addition to traditional elements, it incorporates hardware for sensing, localization, and AI processing. This layer must ensure high data rates, low latency, reliability, energy efficiency, and cost-effectiveness, while remaining adaptable to diverse network topologies for seamless inter-connectivity.
    \item \textit{The Network Services Layer} is envisioned as cloud-native and decentralized. Network functions and services will be software-defined, distributed at the edge, and implemented as modular, containerized microservices. This architecture enhances scalability, flexibility, and responsiveness, enabling service delivery near the RAN for improved latency and user experience.
    \item The \textit{Applications Layer} hosts vertical services, including immersive smart cities, AI-driven transportation, and cyber-physical systems. It also supports native management features such as network slicing, blockchain-based security, sensing integration, and container orchestration.
\end{enumerate}
Figure \ref{6GArchitecture_2} illustrates the high-level architectural model for 6G.
\begin{figure}[h]
    \centering
    \includegraphics[width=\linewidth]{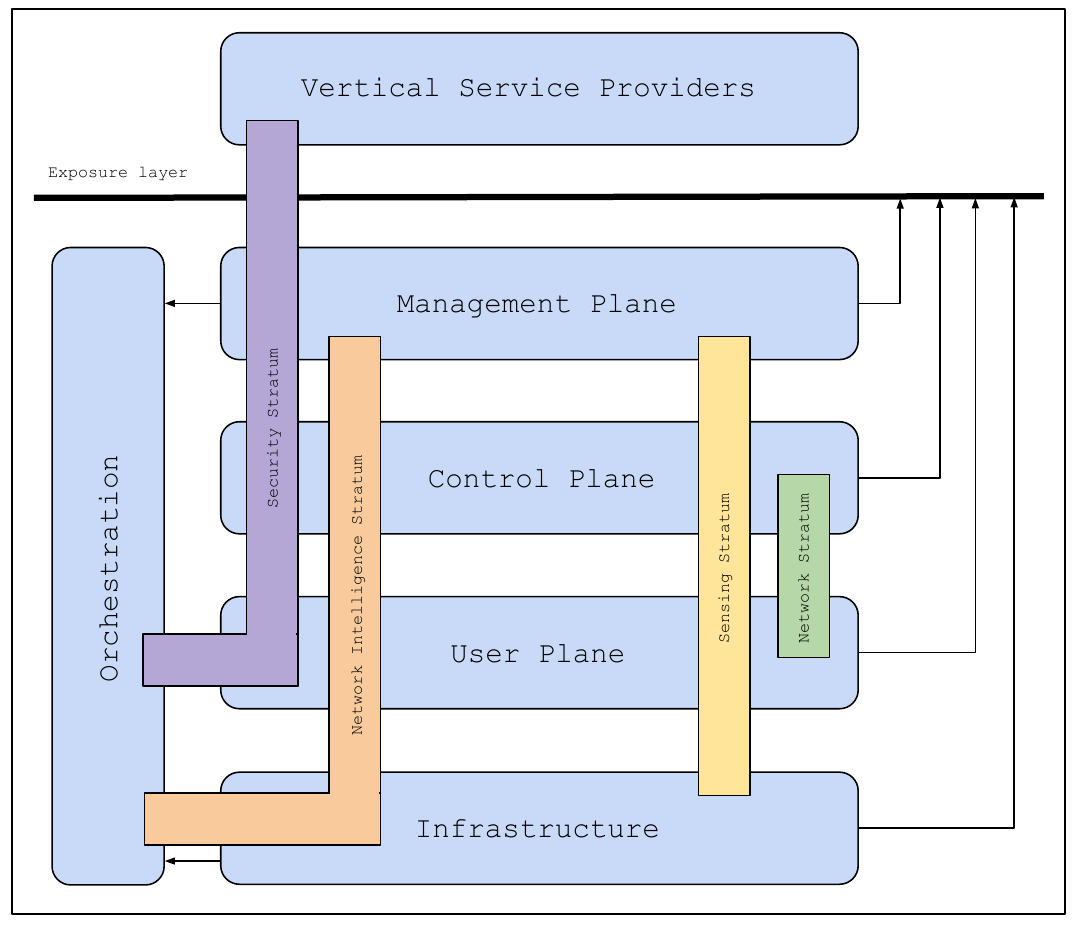}
    \caption{Layers and strata of the 6G architecture illustrating the vertical and horizontal decomposition of functionalities across physical infrastructure, network control, data processing, and application domains. The architecture integrates strata for intelligent management, security enforcement, sensing, and coordinated network access and data exchange.}
    \label{FunctionalLayers}
\end{figure}
%

The architectural functional view is described as a hierarchical structure of lateral planes and transversal strata, similar to prior mobile broadband systems. Planes are logical layers that separates responsibilities based on traffic type, such as user, control, and management data. Strata, in contrast, group planes by protocol functions, including security, sensing, and intelligence. A brief description follows.
\begin{itemize}
    \item The User Plane handles user data (e.g., voice, video, Internet) between user equipment (UE) and the network.
    \item The Control Plane manages signaling tasks like connection setup, session control, and mobility.
    \item The Management Plane oversees network configuration, monitoring, and maintenance to ensure reliability and QoS.
    \item The Orchestration Plane automates resource coordination across network domains and layers, serving as an intelligent extension of the Management Plane for dynamic and complex scenarios.
\end{itemize}
The \textit{Network Stratum} (green) spans the user and control planes, managing network access and enabling data exchange. It must support emerging technologies such as THz communication, RIS, and AI-native air interfaces. The \textit{Intelligence Stratum} (orange) operates across all planes and domains, delivering analytics and automation for intelligent network management. The \textit{Sensing Stratum} (yellow) integrates sensing functions across user, control, and management planes, enabling the network to interact with its physical environment. The \textit{Security Stratum} (purple) covers all planes and domains, ensuring cyber-security and privacy. Unlike the others, it also interfaces with the service provider domain, giving vertical applications enhanced security and control over their data. Figure \ref{FunctionalLayers} shows the layers and strata of the 6G architecture.
\section{Technology Enablers: Innovative Technologies Driving 6G}
\label{TechnologyEnablers}
\begin{figure}
    \centering
    \includegraphics[width=\linewidth]{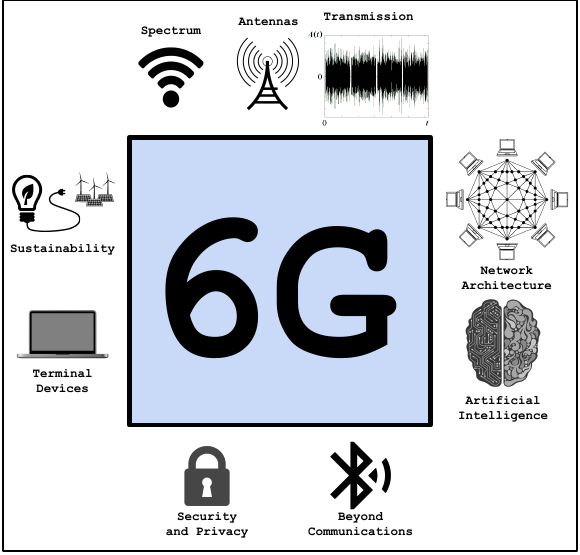}
    \caption{Active research areas for 6G networks highlighting key domains such as spectrum innovations, advanced antenna technologies, transmission techniques, next-generation network architectures, AI-aided communication, beyond-communication services, security and privacy frameworks, terminal device enhancements, and energy-efficient system designs.}
    \label{fig:researchAreas}
\end{figure}

Realizing IMT-2030 will require the convergence of innovations from diverse scientific fields. ITU-R M.2516-0~\cite{itu_r_m2516_2022} provides a broad view of potential technology enablers for the framework, shaped by emerging services, application trends, and relevant driving factors. For clarity, we group technology enablers into nine major research domains: \textit{Spectrum}, \textit{Antennas}, \textit{Transmission}, \textit{Network Architecture}, \textit{Network Intelligence}, \textit{Beyond Communications}, \textit{Security and Privacy}, \textit{Terminal Devices}, and \textit{Sustainability}. Research areas are illustrated in Figure~\ref{fig:researchAreas}. The following subsections outline candidate technologies within each area.
%
\subsection{Spectrum}
\label{Spectrum}
The \textit{Spectrum} research area explores new frequency bands for data transmission~\cite{Matinmikko-Blue2020}, focusing on efficient spectrum use, propagation challenges, resource management, and regulatory frameworks~\cite{mmwave_5g6g}. \textit{THz Communications} operates in the 0.1 to 10 THz range (micrometer wavelengths), offering up to 300 GHz of underutilized bandwidth~\cite{Shafie2023}. Despite limited range and high attenuation, THz enables Tbps data rates, making it suitable for ultra-high-speed, short-range links~\cite{serghiou2022terahertz, rikkinen2020thz}. \textit{OWC} uses light in the visible (430–750 THz), infrared (300 GHz–430 THz), and ultraviolet (750 THz–30 PHz) ranges~\cite{chi2020vlc6g}. While OWC requires line-of-sight and is sensitive to environmental factors like fog and dust, it offers high-capacity free-space transmission, making it a strong candidate for backhaul and low-orbit space links~\cite{arai2021owc6g}. The governance of these new bands is critical to ensure efficient, reliable, and interference-free communication services. Regulatory frameworks must support flexible licensing, intelligent allocation, dynamic aggregation, and fair sharing tailored to 6G demands~\cite{qamar2020spectrum6g, matinmikko_blue2020_spectrum6g, alsaedi2023spectrum6g}.
\subsection{Antennas}
\label{Antennas}
The \textit{Antennas} research area focuses on advanced antenna systems and meta-surfaces to meet 6G performance targets~\cite{ikram2022road6g}. Priorities include ultra-massive antenna arrays, high-frequency operation, energy efficiency, and the integration of intelligent active and passive meta-surfaces for improved signal control~\cite{Hafizah2020}. Ultra-massive MIMO, an evolution of Massive MIMO, uses thousands of antenna elements to enhance spatial resolution, enabling focused energy beams for higher spectral efficiency, ultra-high data rates, and large-scale connectivity~\cite{faisal2020ultramassive}. RIS manipulate signal propagation through programmable meta-surfaces that control wave amplitude, phase, and direction in real-time~\cite{basar2019wirelessRIS}. While RIS increases network complexity, it significantly improves beamforming and coverage in dense environments~\cite{Pan2021}. Despite fabrication and material challenges, RIS is considered foundational for 6G~\cite{basar2019wirelessRIS}. Holographic Radio further advances this concept by shaping wavefronts via boundary surfaces without full channel state information. Holographic RIS enables more precise beamforming and spatial control, making it essential for ultra-dense 6G deployments~\cite{deng2023rhs_holographic}.
\subsection{Transmission}
\textit{Transmission} technologies aim to optimize spectrum usage to meet 6G performance demands. Key research priorities include efficient multiple-access schemes, accurate channel modeling, advanced modulation, and modern coding techniques~\cite{shah2021survey_ma,wang2020channel6g}. Non-Orthogonal Multiple Access (NOMA) allows multiple users to share the same frequency, time, or code resources by assigning different power levels or code sequences~\cite{Liu2022}. Unlike orthogonal methods (e.g., \ac{ofdma}, \ac{tdma}), NOMA overlaps resources, making it well-suited for massive connectivity. Novel Electromagnetic Waveforms are being developed to address limitations at mmWave and THz frequencies, with a focus on power efficiency, low latency, and resilience to interference and Doppler effects~\cite{noor2022oam_review}. Advanced modulation and coding techniques increase spectral efficiency and reliable transmission in noisy and dynamic channels by embedding data in unconventional domains, including higher-order constellations, resource indices, or angular dimensions~\cite{Hadani2017OTFS,Liu2022IMMA}. 
\subsection{Network Architecture}
\label{NetworkArchitecture}
\begin{figure}
    \centering
    \includegraphics[width=\linewidth]{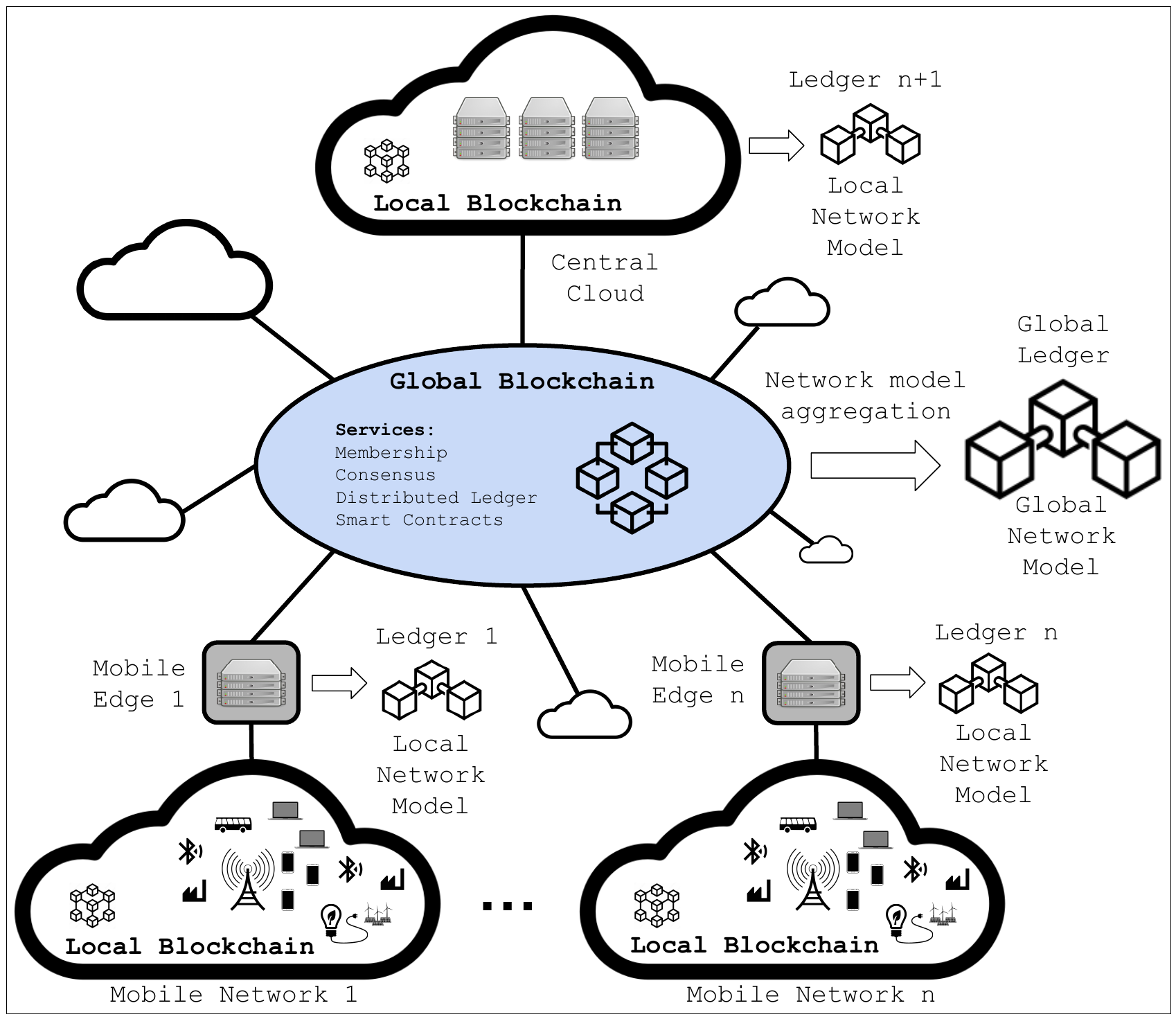}
    \caption{Blockchain-enabled orchestration for 6G networks, illustrating the use of distributed ledger technologies to support secure, transparent, and decentralized management across heterogeneous and multi-domain infrastructures.}
    \label{Figure_6GBlockchain}
\end{figure}

The \textit{Network Architecture} research area focuses on designing the layout, components, and operational principles needed for ultra-fast, intelligent, and large-scale mobile broadband networks~\cite{khan2020_6g_wireless_systems,dogra2021surveybeyond5g}. A key element is network orchestration, which manages resources, functions, and services across a highly heterogeneous architecture, including terrestrial and non-terrestrial infrastructure, virtualized and sliced networks, and distributed (potentially quantum) computing nodes~\cite{shen2022holistic6g,wu2022ai_native_slicing,tang2021computingpowernet}. Traditional centralized management may struggle to scale in 6G due to bottlenecks, single points of failure, and trust limitations among elements like UEs, base stations, and edge nodes~\cite{coronado2022zero_touch}. Research on distributed trust models is essential for building secure communication protocols, node reputation systems, data integrity models, and autonomous decision-making processes. Consortium blockchain models, in particular, may offer lightweight frameworks with membership, consensus, ledger, and smart contract services for efficient and transparent network orchestration~\cite{verarivera2025decentralizingtrustconsortiumblockchains}. Figure \ref{Figure_6GBlockchain} illustrates this concept.
%
\subsection{Network Intelligence} 
\label{NetwrokIntelligence}
AI refers to the ability of machines to perceive, summarize, generalize, and infer from input data, mimicking human cognition~\cite{elsayed2021ai_enabled}. AI methods extract knowledge from data to make predictions and interpret information. The \textit{Network Intelligence} research area explores how to integrate AI into 6G systems to enable adaptive, autonomous, and efficient operations~\cite{letaief2019roadmap6g}. This includes optimizing functions and services based on internal performance metrics and external conditions using efficient, transparent, and distributed AI models and algorithms  ~\cite{du2020ml_6g,she2021urllc_tutorial,sami2021aiscaler,liu2022gan_trust6g,termehchi2025koopmanbasedgeneralizationdeepreinforcement}. Embedding intelligence across the air interface, edge, and cloud layers is essential for managing the system complexity. At the air interface layer, AI can dynamically allocate spectrum, optimize beamforming, and manage interference in real-time~\cite{hoydis2020ai_airinterface6g}. Edge and fog intelligence supports local processing for low-latency decision-making, while cloud intelligence handles large-scale analytics and network coordination~\cite{Letaief2022EdgeAI6G}. Together, these AI-driven layers will form a cohesive system capable of real-time adaptation and continuous optimization.
\subsection{Beyond Communications}
\label{BeyondCommunication}
The \textit{Beyond Communications} research area expands the traditional communication capabilities of cellular networks by integrating sensing, positioning, and environmental monitoring to enable contextual awareness~\cite{Aryan2024}. A key technology is Joint Sensing and Communication (JSAC), which simultaneously transmits data and senses the environment using shared hardware, spectrum, and signaling~\cite{liu2022isac6g}. JSAC can detect spatial layouts~\cite{dong2022s2aas}, monitor motion and trajectories~\cite{xiao2020overview_ilac}, track human posture and gestures, and follow environmental conditions including rain and pollution~\cite{gonzalezprelcic2024isac_revolution}. Combined with AI and distributed computing, JSAC supports real-time processing across cloud, edge, and fog infrastructure~\cite{feng2021jcsc_imtc}, giving 6G networks contextual consciousness. The communication and sensing duality will transform mobile broadband into conscious, actionable systems capable of real-time interaction with their surroundings~\cite{tan2021isac_overview}. JSAC may drive innovation across sectors such as transportation, healthcare, and urban development~\cite{delima2021convergent6g}, enhancing user experience and system functionality. 
\subsection{Security and Privacy}
\label{SecurityAndPrivacy}
Security protects the integrity, confidentiality, and availability of networks and data from unauthorized access, damage, or disruption~\cite{Porambage2021}. Privacy---a subset of security---ensures user data is accessed with explicit consent and protect identities during data collection, transmission, processing, and storage. Research in \textit{Security and Privacy} explores theoretical approaches to build trust in increasingly connected environments~\cite{nguyen2021security6g_survey}. Lightweight cryptography enables encryption, authentication, and key management with minimal overhead, making it ideal for resource-constrained 6G devices. Combined with distributed zero-trust architectures like blockchain~\cite{Khan2021BlockchainIndustry4Survey} can further enhance resilience through continuous verification, least-privilege access, and micro-segmentation. Continued research into scalable and interoperable zero-trust frameworks is essential for secure 6G communications.
\subsection{Terminal Devices}
\label{TerminalDevices}
The \textit{Terminal Devices} research area focuses on advancing technologies for UEs, particularly IoT devices such as sensors, actuators, and meters~\cite{Huo2019Enabling}. These components are essential for collecting, processing, and transmitting data in applications across sectors like healthcare, industrial automation, environmental monitoring, smart cities, and autonomous systems~\cite{Nguyen2022}. 6G devices are expected to be compact, energy-efficient, durable under harsh conditions, and affordable for mass deployments~\cite{Naser2023}. They will combine real-time, high-precision sensing capabilities with advanced control functions that translate digital commands into physical actions, enabling automated machinery and cyber-physical systems~\cite{Lee2020CompliantTEG,Cao2024DielectricGenes,Zhang2023BiocompatibleBiosensors}.
\subsection{Sustainability}
\label{Sustainability}
The \textit{Sustainability} research area focuses on low-emission energy sources, infrastructure, and materials to support  environmentally responsible wireless networks~\cite{huang2020green6g_survey, hu2021energy_innetwork6g}, in alignment with carbon neutrality goals to combat climate change~\cite{un_netzero_2020}. Transitioning to renewable energy---such as solar, wind, hydro, geothermal, and nuclear---is critical to power energy-intensive communication systems and reduce their environmental footprint~\cite{hu2020energy_selfsustain_6g}. Energy-aware protocols with dynamic power management that adapt to real-time traffic can further reduce energy consumption~\cite{mao2021ai_green6g, larsen2023greener_ran_survey, hu2021energy_selfsustain_6g, imoize2022ee_power_udcfmimo}. Sustainability will be a core pillar of 6G that must emphasize energy-efficient design across all network components and layers to meet global sustainability goals.

\section{6G in Action: Unlocking Innovation Opportunities for Society, Businesses, and Industries}
\label{6GVerticals}
\begin{figure}
    \centering
    \includegraphics[width=\linewidth]{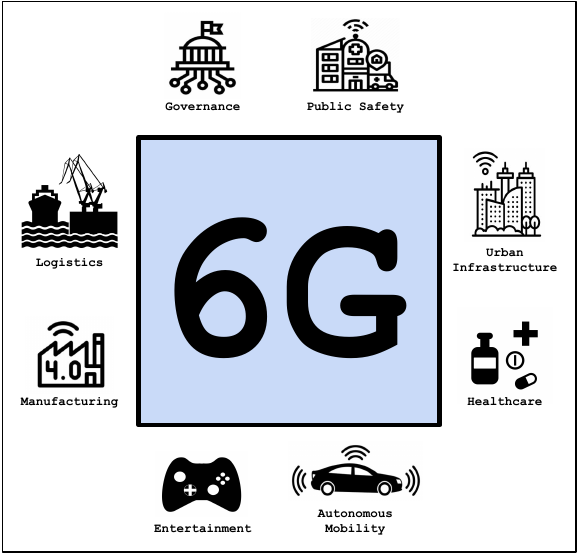}
    \caption{Key vertical domains impacted by 6G-enabled applications, including e-governance, public safety, smart urban infrastructure, personalized healthcare, autonomous mobility, immersive entertainment, smart manufacturing, and autonomous logistics.}
    \label{fig:verticalSectors}
\end{figure}
ITU-R M.2441-0~\cite{itur2018usage} outlines a broad array of telecommunication services
supported by IMT-2030 for emerging applications across numerous vertical sectors. Key vertical sectors are illustrated in Figure~\ref{fig:verticalSectors}. For clarity, we categorize potential 6G vertical applications into three sectors: \textit{Social}, \textit{Business}, and \textit{Industrial}. A discussion follows.
\subsection{Social Sector} 
\begin{table*}[]
\centering
\resizebox{\linewidth}{!}{
\begin{tabular}{|p{2.5cm}|p{7cm}|p{6cm}|p{3cm}|}
\hline
\textbf{Vertical} & \textbf{Candidate Application} & \textbf{Anticipated Innovation} & \textbf{Adoption Barriers} \\
\hline
Governance~\cite{yaacoub2020rural6g_survey, chaoub2022bridging_divide} & 1. Digital government services for citizens, businesses, industries, and public institutions, e.g., voting, taxation, land management, legislation approval, etc. &  Online, fast, secure, and intelligent e-platforms for public service delivery. & Digital divide and outdated regulatory frameworks.  \\
& 2. Digital law enforcement and judiciary applications, e.g., real-time surveillance, predictive policing, judiciary case management, etc. & Online, fast, and intelligent platforms for law enforcement and judiciary processes. & Digital divide and outdated regulatory frameworks. \\
\hline
Public safety~\cite{Hou2022, beniiche2022society5_internet, nguyen2020sdistancing_partii} & 1. Disaster management platforms, e.g., early warning, emergency communication, autonomous response, situational awareness, etc.  & Utra-fast and reliable disaster prevention, response, and recovery systems. & Infrastructure cost and updated regulatory frameworks. \\
& 2. Predictive policing and crime prevention systems, e.g., hot spot analysis, person-based prediction, fraud and cyber-crime prevention, etc. & Crime prediction and prevention using historical data and real-time surveillance analytics. & Public distrust and adequate data privacy regulations. \\
& 3. General surveillance and monitoring systems, e.g., crowd management, user behavior analytics, traffic management, supply chain and employee management, etc. & Real-time observation and analysis of human behavior patterns. & Public distrust and adequate data privacy regulations. \\
\hline
Urban infrastructure~\cite{javed2021future_smart_cities, allam2022_15min_city, murroni2023_6g_smartcity} & 1. Smart utility systems, e.g., smart energy grids, blockchain-enabled energy trading platforms, vehicle-to-grid (V2G) integration platforms, waste-to-energy systems, etc. & Real-time monitoring and management of energy, water, gas, and waste infrastructure. & Infrastructure complexity. \\
& 2. Environmental Monitoring platforms, e.g., greenhouse gas tracking, air quality management, biodiversity and ecosystem monitoring. & Real-time monitoring and management of environmental indicators. & Infrastructure complexity. \\
& 3. Smart Urban Planning tools, e.g., IoT-aided geographic information systems, digital twin-based city modeling tools, city planning platforms for citizens. & Data-driven and intelligent urban planning with citizen-centric design. & Infrastructure complexity. \\
\hline
\end{tabular}    
}
\caption{Potential 6G vertical applications in the social sector, including governance, public safety, and urban infrastructure.}
\label{tab:SocialApplications}
\end{table*}
6G offers transformative potential for social-sector verticals, including governance, public safety, and urban infrastructure~\cite{imoize2021smartinfra6g}. In governance, 6G can power fast, secure, and intelligent e-platforms to improve public services like voting, taxation, land management, law enforcement, and judiciary functions. For public safety, it enables real-time disaster response, early warning systems, emergency communication, and autonomous recovery. Advanced tools like behavioral analytics, fraud detection, and predictive policing may also become viable. In urban infrastructure, 6G's support for massive sensor networks allows real-time monitoring of cities and utilities. Its AI-native and blockchain-based framework will facilitate platforms for digital twin urban planning and decentralized energy trading, including vehicle-to-grid (V2G) and home-to-grid (H2G) systems. Despite challenges, such as public trust, infrastructure costs, and regulatory readiness, 6G can significantly improve public services, safety, and sustainable urban development. Table~\ref{tab:SocialApplications} summarizes applications in the social sector.
\subsection{Business Sector} 
\begin{table*}[]
\centering
\resizebox{\linewidth}{!}{
\begin{tabular}{|p{2.5cm}|p{7cm}|p{6cm}|p{3cm}|}
\hline
\textbf{Vertical} & \textbf{Candidate Application} & \textbf{Anticipated Innovation} & \textbf{Adoption Barriers} \\
\hline
Healthcare~\cite{nayak2021vision6g_healthcare, nguyen2023intelligent_healthcare_iot, hadi2020patientcentric_hetnets} & 1. Remote diagnostics and tele-medicine platforms, e.g., virtual consultation systems, medical imaging and diagnosis, wearable health monitoring devices and networks. & Real-time, accurate, and location-independent medical consultation and diagnosis. & Public distrust. \\
& 2. Personalized medicine platforms, e.g., genomics and precision therapy, AI-assisted drug development, digital twin-aided drug prescription, smart implants and biosensors, etc. & Real-time, hyper-personalized medical analysis, diagnosis, treatment, and prescriptions. & Infrastructure complexity and public distrust. \\
& 3. Remote surgery networks, e.g., AI- and robotics-assisted surgical systems.  & Remote access to specialized surgical experts. & Infrastructure complexity and public distrust. \\
\hline
Autonomous Mobility~\cite{He2021, noor2022_6g_v2x, pandey2024_recent_survey_6g_vehicular} & 1. Smart personal mobility, e.g., autonomous and semi-autonomous cars, IoT-enabled smart scooters and bikes, urban micro-mobility, and mobility-as-a-service (MaaS). & Safe, efficient, personalized, and cost-effective personal transportation systems. & Infrastructure gaps, high initial costs, interoperability challenges, and lack of regulatory frameworks. \\
& 2. Smart public transportation systems, e.g., real-time fleet management, autonomous buses and trains, and multi-modal transportation integration. & Safe, efficient, and cost-effective public mobility solutions. & Infrastructure gaps, high initial costs, interoperability challenges, and lack of regulatory frameworks. \\
\hline
Entertainment~\cite{Zawish2024, rostami2022metaverse_multiverse_wearables} & 1. Immersive content platforms, e.g., gaming, remote work and collaboration, education and training, e-commerce and retail, social media and networking, and virtual events. & Content-rich platforms delivering immersive (e.g., virtual, augmented, or mixed reality) and haptic user experiences. & Infrastructure complexity. \\
\hline
\end{tabular}    
}
\caption{Potential 6G vertical applications in the business sector, including healthcare, autonomous mobility, and entertainment.}
\label{tab:BusinessApplications}
\end{table*}
In business-sector verticals, 6G presents significant opportunities in healthcare, autonomous mobility, and entertainment. In healthcare, it can support reliable remote diagnosis and tele-medicine services, including virtual consultations, medical imaging, remote surgery, and real-time health monitoring. 6G's native intelligence may also facilitate drug development, digital twin-based prescriptions, and smart implants and biosensors for personalized treatment. In autonomous mobility, 6G can provide ultra-fast communication and precise positioning for self-driving vehicles, as well as IoT-enabled micro-mobility solutions like scooters and bikes. This enables safer and more efficient transportation systems with integrated multi-modal mobility. In entertainment, it will deliver immersive experiences for gaming, remote work, education, and social media and live-events. Near real-time responsiveness and native intelligence will support interactive VR, AR, MR, and haptic applications. Despite the benefits, infrastructure complexity, IoT interoperability, and regulatory gaps (particularly in the healthcare sector) remain important adoption limitations still unresolved. Table~\ref{tab:BusinessApplications} summarizes the main 6G applications in the business sector.
\subsection{Industrial Sector}
\begin{table*}[]
\centering
\resizebox{\linewidth}{!}{
\begin{tabular}{|p{2.5cm}|p{7cm}|p{6cm}|p{3cm}|}
\hline
\textbf{Vertical} & \textbf{Candidate Application} & \textbf{Anticipated Innovation} & \textbf{Adoption Barriers} \\
\hline
Manufacturing~\cite{maddikunta2022industry, adel2022_industry5_future, humayun2021industrialrevolution5_cet} & 1. Smart manufacturing applications, e.g., real-time monitoring and process optimization, predictive maintenance, digital twin-based planning, 3D printing.  &  Real-time, data-driven management of manufacturing operations.  & Infrastructure complexity, high initial costs, and labor displacement concerns. \\
& 2. Robot-assisted manufacturing, logistics, and production, e.g., robot-based assembly, material handling, manufacturing, quality inspection, and worker safety training. &  Unsupervised operations and improved productivity, quality, and cost-efficiency. & Infrastructure complexity, high initial costs, and labor displacement concerns.\\
\hline
Logistics~\cite{chen2021wireless_multirobot_smartfactories, padhi2021iiot6g} & 1. Autonomous freight applications, e.g., autonomous freight transport, automated warehousing, etc. & Safe, efficient, and globally scalable freight solutions for manufacturing and delivery. & Infrastructure complexity, high initial costs, and labor displacement concerns. \\
\hline
Agriculture~\cite{polymeni2023_6g_iot_agri5, ranjha2022uav_urllc_agri4} & 1. Precision farming applications, e.g., soil and crop monitoring, pest and disease control, yield optimization, climate-adaptive practices, circular agriculture, etc. &  Real-time and intelligent management. & Labor displacement concerns. \\
& 2. Agricultural drones and robotics applications, e.g., precision spraying, irrigation monitoring, autonomous planting, robotic harvesting, greenhouse automation, & Safe, efficient, and autonomous agricultural operations. & Infrastructure complexity, high initial costs, and labor displacement concerns \\
\hline
\end{tabular}    
}
\caption{Potential 6G vertical applications in the industrial sector, including manufacturing, logistics, and agriculture.}
\label{tab:IndustrialApplications}
\end{table*}
6G's global coverage, AI-native design, and sensing capabilities can transform industrial sectors such as manufacturing, logistics, and agriculture. In manufacturing, it can enable unsupervised production, digital twin planning, real-time monitoring, and predictive maintenance. It may also power 3D printing and robot-assisted tasks for material handling, assembly, and quality control. In logistics, 6G will support autonomous freight transport, global tracking, predictive maintenance, dynamic routing, and seamless cross-border operations. Smart warehousing will benefit from real-time inventory tracking, robotic handling, and integrated supply chains. In agriculture, it will facilitate precision farming for soil and crop monitoring, pest control, yield optimization, and climate-adaptive practices. Drones and robots can automate irrigation, planting, harvesting, and greenhouse management. Overall, 6G will enable autonomous, intelligent, and efficient industrial operations, enhancing aggregate productivity. However, challenges such as infrastructure complexity, high implementation costs, and potential labor displacement must be addressed to ensure adoption success. Table~\ref{tab:IndustrialApplications} summarizes potential 6G applications in industry.
\section{6G Research Frontier: Current Challenges and Future Directions}
\label{6GFrontier}
As 6G moves from conceptualization to implementation, numerous and complex research challenges remain unresolved, spanning physical-layer design, architectural innovation, infrastructure scalability, and barriers for achieving sustainability targets.  

THz communication and OWC face major limitations due to high attenuation, limiting coverage and data rate capacity. Accurate channel modeling and beamforming techniques are critical to mitigating these effects. Ultra-massive MIMO and RIS-assisted beamforming may help meet data rate demands, but, high-frequency operation also requires compact, high-precision antenna arrays, which can be difficult and costly to fabricate. Future solutions may involve multi-band, self-configurable nano-antennas using novel materials---such as meta-materials and graphene—for near-lossless transmission with reduced construction complexity~\cite{Poulakis2022}. NOMA is a promising multiple-access technique that allows several devices to share the same channel with minimal interference by assigning different power levels or codes. However, as the number of users grows, decoding and interference management become increasingly complex. Advanced modulation and coding schemes will be essential to overcome these challenges and ensure scalability in future mobile networks.

While NTNs are essential for achieving global coverage, beyond 6G systems may consider extending connectivity further to enable interstellar communication. This will require advanced satellite systems integrated with possibly quantum computing and communications to power high capacity data transfer~\cite{Wang2022}. As mobile networks grow in scale and complexity, efficient orchestration is essential for coordination and trust. Centralized management is increasingly problematic due to scalability, latency, reliability, and security limitations~\cite{verarivera2025decentralizingtrustconsortiumblockchains}. A more viable solution is decentralized network orchestration, which distributes workloads, functions, and services across cloud, edge, fog, and air interface layers. Blockchain-based orchestration, supported by lightweight cryptographic and consensus protocols, offers a scalable and secure approach for managing the size and heterogeneity of future mobile broadband systems.

For the first time, mobile broadband will embed AI as a core system component. 6G will distribute intelligence across the air interface, edge, and core levels, enabling autonomous operation, real-time optimization, and seamless end-user application support. However, integrating AI presents several challenges. Most algorithms demand high computational power and energy, limiting scalability---particularly on energy-constrained IoT devices. Advancing carbon-neutral energy sources and efficient energy management is critical for enabling AI-native mobile broadband ecosystems. Moreover, many AI models, especially deep learning, operate as "black boxes," making their decisions hard to interpret. Research in explainable (XAI)~\cite{Guo2020} and transparent AI is essential to building user and operator trust. Environmental sensing will also be central to 6G and beyond, enabling contextual awareness for intelligent functions and services. Its integration with communication demands extensive research on joint waveform design, interference-aware resource allocation, and signal processing to extract sensing data from communication signals.

Future research in mobile broadband will span a wide array of technologies to meet evolving societal and industrial demands. Key areas include new carrier frequencies, meta-material antennas, efficient modulation and coding, decentralized orchestration, further AI integration, zero-energy sensors, lightweight cryptography, and carbon-neutral infrastructure. Advancing beyond the 6G frontier will require sustained efforts to drive innovation and unlock the potential of future mobile broadband systems. Table~\ref{tab:IndustrialApplications} summarizes 6G's research frontier.
\begin{table*}[]
\centering
\resizebox{\linewidth}{!}{
\begin{tabular}{|l|l|p{6cm}|p{6cm}|}
\hline
\textbf{Research Area} & \textbf{Technology Enabler} & \textbf{Research Frontier} & \textbf{Future Direction} \\
\hline
Spectrum &  THz Communications \& OWC & Severe propagation limitations, e.g., high path loss, limited difsraction, strong scattering, high absorption, and poor difrraction. & Accurate channel models and advanced beamforming techniques to mitigate propagation challenges.  \\
\hline
Antennas & MIMO \& RIS & High fabrication complexity of compact, high-performance antenna arrays. & Matematerials- and graphene-based nano-antennas to enhance performance and reduce construction complexity.   \\
\hline
Transmission & NOMA \& variations & Decoding and interference management challenges in large-scale NOMA networks. & Advanced modulation and coding schemes to ensure scalable and efficient multiple access. \\
\hline
Network Architecture & NTNs & Highly heterogeneous and complex NTN architectures requiring efficient integration with terrestrial networks and coordinated management. & Lightweight and scalable trust models for decentralized integration and orchestration. \\
& Distributed Computing & Scalability challenges due to the high computation demands of AI-native 6G networks & Distributed cloud, edge, and fog computing models integrated across all network layers and domains. \\
\hline
Network Intelligence & AI models & Transparency concerns due to the "black box" nature of most AI systems. & Explainable and generalizable AI models for transparent and trustworthy decision-making. \\
\hline
Beyond Communications & JSAC & Limited sensing resolution, interference management, and integration with existing network protocols. & Advanced waveform designs for interference-aware transmission and sensing resolution. \\
\hline
Security and Privacy & Blockchain-enabled Security & High-computational overhead from cryptography and consensus mechanisms. & Lightweight cryptographic and consensus protocols tailored for resource-constrained 6G environments. \\
\hline
Terminal Devices & Smart UEs, wearables, and sensors & Integration of communication, sensing, actuation, and computation in compact devices. & Nano-electronic components and metamateials for multifunctional UEs, wearables, and sensors. \\
\hline
Sustainability & Green Communication Protocols & Trade-offs between carbon neutrality and system performance. & Energy-aware protocols
with dynamic power management that adapt to QoS/QoE. \\
\hline
\end{tabular}    
}
\caption{Summary of research challenges and potential future directions for major 6G research domains.}
\label{tab:SocialApplications}
\end{table*}
\section{Lessons Learned}
Despite market challenges, investment in cellular technologies continues, driven by their potential to transform vertical sectors, create new value streams, and address emerging societal and industrial needs~\cite{McKensey2024}. Developing technologies for IMT-2030 requires a sustained intellectual and financial commitment, involving close collaboration among academia, industry, and SDOs. 3GPP anticipates a seven-year timeline from the framework publication in 2023 to the first set of specifications by 2030. Standardization will begin with a requirements study in Release 19, followed by work item reviews in Release 20, and implementable standards in Release 21. Further advancements are expected in Releases 22 and 23. In the coming years, 6G will transition steadily from research efforts to real-world adoption.

Industries may face adoption barriers such as financial constraints, unclear return on investment, and interoperability issues with legacy infrastructure. The varying maturity of enabling technologies will also influence the pace of deployment. For example, quantum computing and communications remain in early research, while RIS, holographic radio, immersive reality, and renewable energy technologies are experimental, with limited pilots. In contrast, technologies like NTNs, classical distributed computing, AI, and cyber-security are rapidly maturing, with commercially available solutions. As these technologies evolve, we expect broader 6G adoption will follow in alignment with the release of the first set of standards around 2030, mirroring the case of previous cellular generations.

Communities without access to communication technologies have limited access to information, public services, and economic opportunities, creating a digital gap that restricts their aspirations for prosperity. The vision for global connectivity seeks to bridge this divide by offering widespread and affordable communication services, promoting digital inclusion. This is crucial for enabling new business and industry models across vertical segments. Sectors such as healthcare, transportation, entertainment, energy, manufacturing, logistics, and agriculture can redefine revenue streams and enhance services through 6G-enabled applications. While the system offers transformative potential, realizing its full value will require supportive government policies, a skilled workforce, public trust, and greater awareness of its benefits and opportunities. 
\section{Conclusions}
\label{Conclusions}
IMT-2030 provides the conceptual framework for 6G cellular networks. 6G is set to redefine mobile broadband by moving beyond the \textit{``more bandwidth, less latency"} conversation. Academia, industry players, regulators, and standardization bodies are already engaged in early discussions intended to shape future technical standards. This review maps the 6G landscape through a thematic synthesis aligned with IMT-2030, providing insights for specialists and generalists involved in standardization efforts and broader 6G research.

We identify five core themes mirroring the natural evolution from 5G to 6G. The themes are: (1) 6G vision, (2) Use Cases, Performance Requirements, and Architectural Trends (3) Enabling Technologies, (4) Impact on Vertical Sectors, and (5) the 6G Research Frontier. The majority of the screened literature (69.9\%) focus on technology enablers, reflecting strong research interest in foundational technologies for the system. In contrast, vertical applications receive less attention (19.8\%), despite their importance. A smaller share of the contributions corresponds to 6G general surveys (8.5\%), and the research frontier for the system (0.9\%). Within the enabler-focused literature (628 articles), key areas include antennas and meta-surfaces (20.22\%), terrestrial and non-terrestrial network integration (14.49\%), AI-based architectures (13.54\%), and high-frequency transmission technologies (13.22\%). While these technologies are indeed essential for the system, other critical areas like network orchestration (4.78\%) and security and privacy (4.46\%) appear underexplored despite their equal relevance.

In summary, this review highlights the research focus across IMT-2030 thematic domains. By mapping the current landscape, we aim to support 6G stakeholders in discussions, collaboration, and funding decisions for future standards-informed research, particularly in overlooked enabler and application domains within the framework. 

\bibliographystyle{IEEEtran}
\bibliography{references}

\begingroup
\let\description\LaTeXdescription
\let\enddescription\endLaTeXdescription
\endgroup
\begin{IEEEbiography}[{\includegraphics[width=1in,height=1.25in,clip,keepaspectratio]{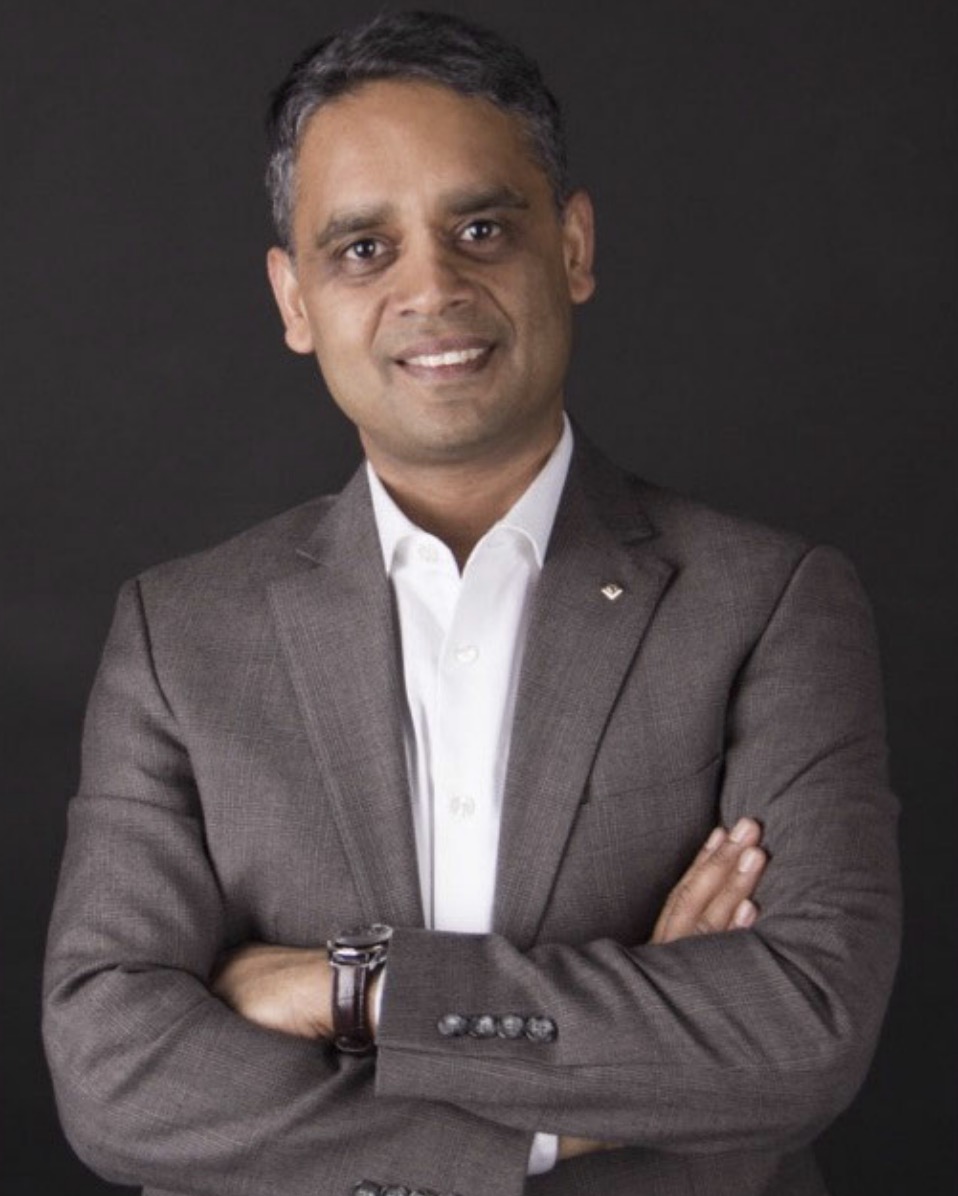}}]{Ekram Hossain}(Fellow, IEEE) is a Professor and the Associate Head (Graduate Studies) of the Department of Electrical and Computer Engineering, University of Manitoba, Canada. He is a Member (Class of 2016) of the College of the Royal Society of Canada. He is also a Fellow of the Canadian Academy of Engineering and the Engineering Institute of Canada. He has won several research awards, including the 2017 IEEE Communications Society Best Survey Paper Award and the 2011 IEEE Communications Society Fred Ellersick Prize Paper Award. He was listed as a Clarivate Analytics Highly Cited Researcher in Computer Science in 2017-2024. Previously, he served as the Editor-in-Chief (EiC) for the IEEE Press (2018–2021) and the IEEE Communications Surveys and Tutorials (2012–2016). He was a Distinguished Lecturer of the IEEE Communications Society and the IEEE Vehicular Technology Society. He served as the Director of Magazines (2020-2021) and the Director of Online Content (2022-2023) for the IEEE Communications Society.
\end{IEEEbiography}
\begin{IEEEbiography}[{\includegraphics[width=1.1in,height=1.10in,clip, keepaspectratio]{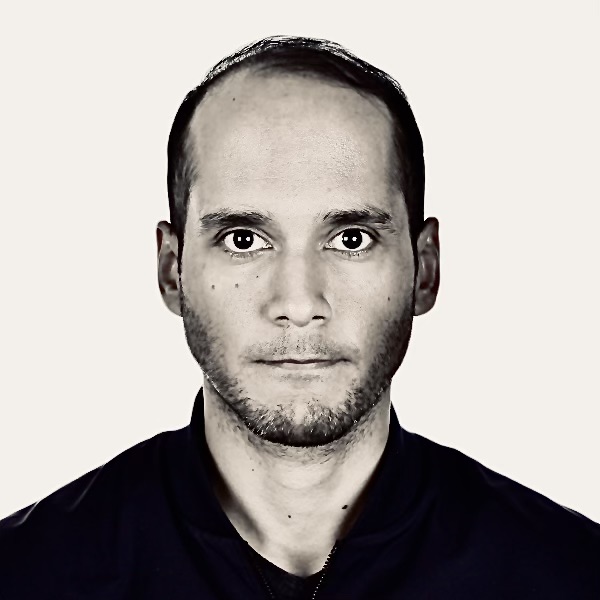}}]
{Angelo Vera-Rivera} is a Research Technician in the Department of Electrical and Computer Engineering at the University of Manitoba. He works at the Wireless Communications, Networks, and Services Laboratory (Wicons Lab) under the supervision of Professor Ekram Hossain. Angelo holds an M.Sc. degree in Electrical and Computer Engineering from the University of Manitoba (Canada, 2022), an M.Sc. in Telecommunications from George Mason University (United States, 2015), and a B.Sc. in Electronics and Telecommunications from Escuela Superior Politecnica del Litoral (Ecuador, 2011). He is a member of Engineers Geoscientists Manitoba (EIT) and the IEEE (Professional Member). His research interests focus on the intersection of Blockchain and Edge Computing technologies with mobile communication systems. 
\end{IEEEbiography}
\end{document}